# Radio Astronomy and eVLBI using KAREN

S.D.Weston*, T. Natusch and S. Gulyaev

Institute for Radio Astronomy and Space Research

AUT Univeristy

Private Bag 92006

Auckland 1142

New Zealand

*stuart.weston@aut.ac.nz

**Abstract:** Kiwi Advanced Research and Education Network (KAREN) has been used to transfer large volumes of radio astronomical data between the AUT Radio Astronomical Observatory at Warkworth, New Zealand and the international organisations with which we are collaborating and conducting observations. Here we report on the current status of connectivity and on the results of testing different data transfer protocols. We investigate new UDP protocols such as "tsunami" and UDT and demonstrate that the UDT protocol is more efficient than "tsunami" and ftp. We report on our initial steps towards real-time eVLBI and the attempt to directly stream data from the radio telescope receiving system to the correlation centre without intermediate buffering/recording.

## 1 INTRODUCTION

The New Zealand 12-m radio telescope is located some 60 km north of the city of Auckland, near the township of Warkworth as shown in Figure 1. Currently it operates in three frequency bands centred around 1.4, 2.4 and 8.4 GHz. This fast slewing antenna (5 degrees per second in Azimuth) is well suited to the purposes of geodetic radio astronomy and spacecraft navigation/tracking [1]. It also can be effectively used for astrophysical VLBI (Very Long Baseline Interferometry) research in conjunction with large radio telescopes [2].

VLBI is a technique that uses several radio telescopes separated by continental distances, synchronised and working together, simultaneously tracking the same celestial object, therefore creating a virtual radio telescope of a continental/trans-continental scale. This method was successfully used in April-May 2010 when a combination of six radio telescopes in Australia and New Zealand (Warkworth) observed the core of the radio galaxy Centaurus A [3]. Following the installation of the KAREN (Kiwi Advanced Research and Education Network) [4] GigaPoP at the AUT radio telescope, the New Zealand Radio Telecope data was transferred to Western Australia, where it was correlated and calibrated, and images of a sub-parsec resolution were created.

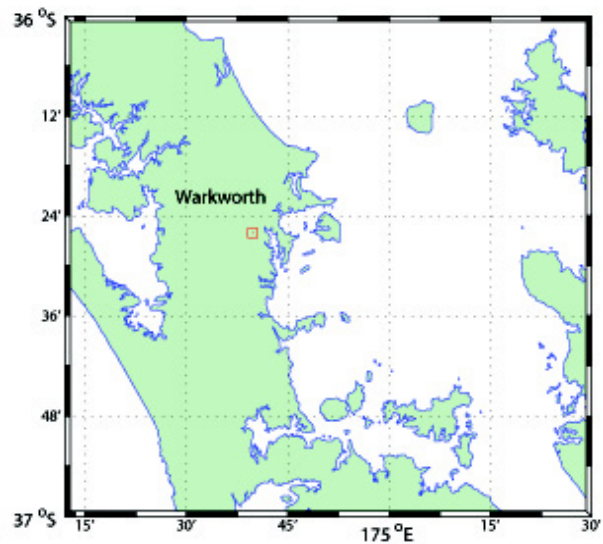

Figure 1. □ Shows geographic location of the AUT Radio Telescope south of Warkworth.

Figure 2 shows a screenshot of KAREN Weathermap for Centaurus A data transfer between Warkworth and Sydney on 9-13 May 2010. Data transfer is an issue of great importance for VLBI. Previously data was recorded to magnetic tapes, these were then sent via traditional means such as post/courier to the correlation centre. More recently with the reduced cost of hard disk storage, the data has been recorded to removable disk arrays which can then also be sent physically to the correlation centre. Recently with the advent of eVLBI the data is sent via high speed networks to the correlation centres.

With the connection of KAREN to the AUT Radio Telescope we wish to optimize the use of KAREN for transferring large volumes of observational data to our partners in Australia, the United States and Europe and for conducting eVLBI. Here we discuss the use of FTP over TCP/IP for sending these files to Australia, as well as new protocols which are being used by our partners such as tsunami [5] and UDT [6] (UDP-based Data Transfer) via the network protocol UDP.

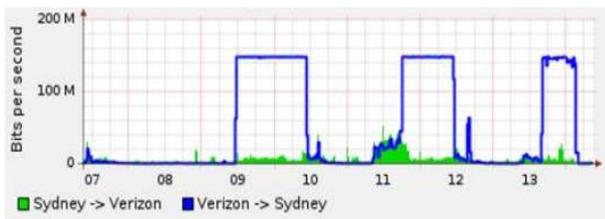

Figure 2. KAREN Weathermap data transfer between Warkworth and Sydney on 9-13 May 2010 (VLBI observations of Centaurus A radio galaxy) using ftp.

## 2 BACKGROUND

Point to point with no hops FTP is efficient, but as the number of hops in the route increases and the incidence of lost packets and collisions increases the TCP congestion avoidance algorithm becomes a severe limitation to the throughput that can be achieved.

*2.1 Overview*

The Institute for Radio Astronomy and Space Research collaborates with a number of international partners. This research collaboration can be conditionally broken into three major research topics/groups:

1. Astrophysical VLBI observations with Australian Long Baseline Array (LBA), data from the AUT Radio Telescope has to be sent to CSIRO [7] and then to Curtin University, Perth for correlation and imaging. The main goal of this collaboration is the study of the physics and origin of both extragalactic and galactic radio sources (active galactic nuclei, radio galaxies, supernova remnants, star formation regions and others).

2. Observation and navigation of inter-planetary missions/spacecraft, as well as ground tracking services for a variety of space missions. VLBI observations of Japan Aerospace Exploration Agency (JAXA) [8] IKAROS and Akatsuki spacecraft were conducted in July 2010. In September 2010 a signal from European Space Agency (ESA) [9] Mars Express was successfully detected. Using a PCEVN recorder and a modified receiver backend a 32 MHz band of data was collected for analysis. All data was transferred by network directly to Metsähovi, Finland [10] and to the Joint Institute for VLBI in Europe (JIVE) [11] using the international connections of KAREN.

3. The forthcoming regular IVS (International VLBI Service for Geodesy and Astrometry) [12] observations of a range of quasars, which uniquely support the International Celestial Reference Frame. We will be required to send observational data to one of two correlation centres located in the United States Naval Observatory and the Max-Planck Institute for Radio Astronomy in Germany.

Figure 3 gives a simple visual overview of destinations we are trying to establish connectivity with via KAREN for current and future data transfers from the Warkworth Radio Astronomical Observatory site. The double ended arrow lines are solid to indicate that connectivity has been established between the Observatory and the remote centres. The dashed line shows a connection needed but not yet established.

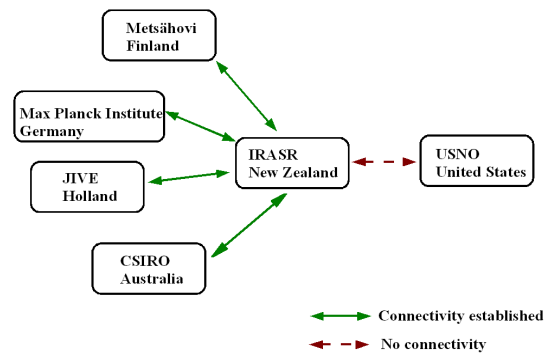

Figure 3: Established and required connectivity from IRASR to partners via KAREN

*2.2 Network Protocols*

The first UDP file transfer protocol we investigated was tsunami developed by Jan Wagner of the Metshovi Radio Observatory in 2007 [5]. This will be the protocol of choice for sending files to Bonn for the coming IVS observations, as stipulated by the Max-Planck Institute.

Another UDP protocol called UDT was developed in the University of Illinois in 2005 [13]. UDT was investigated by our Australian partners (private communication Chris Phillips, CSIRO) in 2008-09. This now appears to have matured and is of further interest and warrants further investigation.

*2.3 Connectivity Status*

Table 1 gives an overview of destinations that connectivity has been achieved with and the protocols that have been verified for data transfer and command line access to remote servers for initiating data transfers.

| Destination | Protocol | Command | | Date |
|---|---|---|---|---|
| CSIRO | ftp | | ssh | 01/04/10 |
| Bonn | UDT | tsunami | ssh | 01/06/10 |
| Metsähovi | UDT | tsunami | ssh | 21/07/10 |
| JIVE | UDP | | iperf | 27/07/10 |
| USNO | - | - | - | - |

Table 1: Connection details between AUT and partners via KAREN

# 3 DATA TRANSFER TESTS and RESULTS

*3.1 Max-Planck Institute, Bonn, Germany*

The results in Table 2 were obtained in June 2010 transferring an actual VLBI file (16 bit) produced by an observation using the AUT Radio Telescope. The data was transferred from the AUT's IBM Blade server via the KAREN network to the Max-Planck Institute, Bonn, Germany. The second column shows the protocol used, third column gives the amount of data sent in bytes, the forth column provides the time it took to transfer the data and the fifth column is the result of the division of data volume by time, therefore providing an average throughput rate over the period.

| Route | Protocol | Bytes | Time (s) | Throughput (Mbps) |
|---|---|---|---|---|
| AUT – Bonn | Ftp | 65G | 8016 | 65 |
| AUT – Bonn | Tsunami | 65G | 3466 | 151 |
| AUT – Bonn | UDT | 65G | 1920 | 273 |

Table 2: Data Transfer results AUT to Bonn

This exercise was conducted repeatedly over several days and at different times giving slightly different average rates.

The following is the traceroute from the AUT Blade to the IP address given for Bonn. This provides a snapshot in time for a path to the destination which can change.

```
[evlbi@ska-blade1 vt14b]$ traceroute viola2

traceroute to viola2 (194.94.199.165), 30 hops max, 40 byte packets

1 156.62.231.65 (156.62.231.65) 0.478 ms 0.531 ms 0.609 ms

2 210.7.32.1 (210.7.32.1) 0.368 ms 0.372 ms 0.438 ms

3 210.7.36.67 (210.7.36.67) 0.663 ms 0.773 ms 0.823 ms

4 210.7.47.22 (210.7.47.22) 143.304 ms 143.417 ms 143.468 ms

5 abilene-1-lo-jmb-706.sttlwa.pacificwave.net (207.231.240.8) 143.210 ms 143.216 ms 143.239 ms

6 xe-2-0-0.0.rtr.salt.net.internet2.edu (64.57.28.104) 159.758 ms 159.704 ms 159.694 ms

7 xe-1-1-0.0.rtr.kans.net.internet2.edu (64.57.28.25) 186.049 ms 184.231 ms 190.001 ms

8 ge-1-2-0.0.rtr.chic.net.internet2.edu (64.57.28.37) 200.868 ms 200.927 ms 200.921 ms

9 xe-3-0-0.0.rtr.wash.net.internet2.edu (64.57.28.101) 217.103 ms 217.101 ms 217.143 ms

10 abilene-wash.rt1.fra.de.geant2.net (62.40.125.17) 310.233 ms 310.269 ms 310.307 ms

11 dfn-gw.rt1.fra.de.geant2.net (62.40.124.34) 310.920 ms 310.932 ms 310.927 ms

12 xr-bir1-te1-1.x-win.dfn.de (188.1.145.45) 313.058 ms 313.165 ms 313.403 ms

13 xr-bon1-te2-1.x-win.dfn.de (188.1.144.10) 307.749 ms 308.166 ms 307.499 ms

14 viola2 (194.94.199.165) 308.847 ms 308.850 ms 308.826 ms
```

It has been found that issuing this command repeatedly over several months the route appears static and has not changed. The total number of hops is 14 on the route from AUT to Bonn via KAREN, demonstrating the complex path to be negotiated and why data transfers via protocols such as FTP will not be efficient.

*3.2 Metsähovi Observatory, Finland*

The results in Table 3 were obtained in July 2010 transferring an actual VLBI file (16 bit) produced by the AUT Radio Telescope during our VLBI observations. The data was transferred from the AUT's IBM Blade server via the KAREN network to the Metsähovi Observatory, Finland. The layout and description of Table 3 is the same as for Table 2 for easy comparison.

| Route | Protocol | Bytes | Time (s) | Throughput (Mbps) |
|---|---|---|---|---|
| AUT – Metsähovi | Ftp | 3.1G | 7458 | 61 |
| AUT – Metsähovi | Tsunami | 65G | 4979 | 105 |
| AUT – Metsähovi | UDT | 65G | 1157 | 453 |

Table 3: Data Transfer results AUT to Metsähovi

At the end of August 2010 observations were conducted of the ESA's Mars Express spacecraft orbiting Mars. This data was transmitted to Metsähovi via KAREN. The data of 24 August which totaled 86 GB was sent using the UDP protocol tsunami immediately after the observational session (Figure 4). The next set of observational data (26 August) totaling 187 Gbytes was sent via UDT (Figure 5). Average rates sustained were 250 and 300 Mbps respectively.

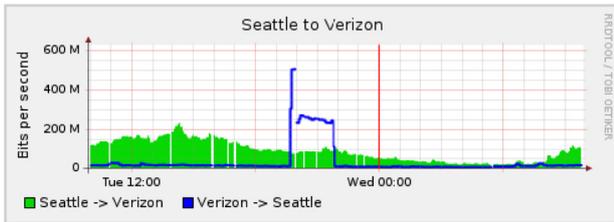

Figure 4: Snapshot of KAREN weathermap 24/8/2010 for period showing file transfers to Metsähovi using tsunami

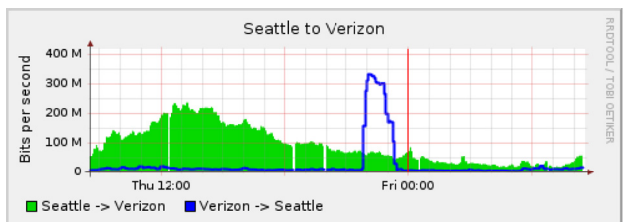

Figure 5: Snapshot of KAREN weathermap 26/8/2010 for period showing file transfers to Metsähovi using UDT

It was found that UDT performed better than tsunami by transmitting more data in a shorter period and obtaining a higher bandwidth utilization during that period.

### 3.3 Comparison of results for file transfer

The main purpose of our tests was to compare the performance of different data transfer protocols. Figure 6 shows the result of this comparison for two different end points, namely Bonn and Metsähovi. The histogram clearly demonstrates the advantage of the UDT protocol. It is up to 4 times faster than tsunami protocol and almost one order of magnitude faster than the standard ftp. The difference in data transfer rate between end points can be connected with the varying number of hops and performance of sections along the routes.

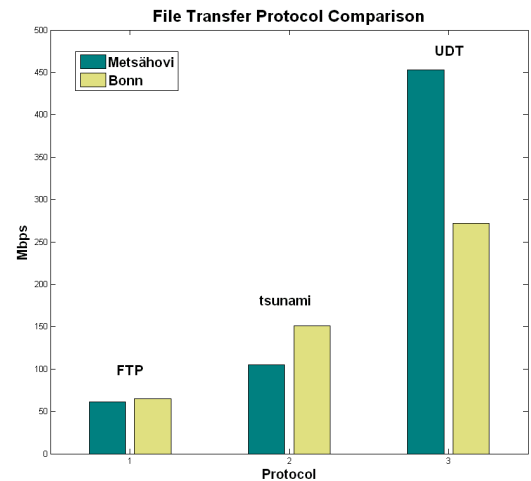

Figure 6: Comparison of the different file transfer protocols to different network end points over KAREN

### 3.4 Data streaming results

Another set of experiments was conducted in September 2010 aiming to test the tsunami protocol for streaming VLBI data directly from the radio telescope receiving system via KAREN to Metsähovi. This test is an important step towards real-time eVLBI.

Initial tests from the Observatory at Warkworth to the IBM Blade server at the AUT City Campus showed that the tsunami rates of 485 Mbps with no lost packets was sustainable. However, when streaming data from Warkworth to Metsähovi many thousands of lost packets occurred and a sustainable rate of 350 Mbps was achieved. This is significantly lower than the rate of 512Mbps required for the real-time eVLBI streaming of 8bit data to Metsähovi.

The next step we are planning to undertake is to test the more efficient UDT protocol for direct data streaming. It requires the VLBI software to be modified to make use of UDT instead of tsunami for streaming. This will require some programming effort as the network code has been merged within the application code rather than being maintained in a separate library.

## 4 CONCLUSIONS

It was clearly demonstrated that the use of the UDT protocol for radio astronomical data transfer has a number of advantages compared to the protocols currently used for VLBI and eVLBI. In particular, UDT has some advantages over tsunami:

• UDT is a better citizen on the network leaving bandwidth for TCP and other UDP protocols, which is very important on a shared network such as KAREN.

• UDT has an application programming interface (API) allowing easy integration with existing or future applications.

We have also investigated the ability to stream collected data via UDT and have modified the Curtin 16 bit to 2 bit conversion program for data streaming to a remote server in a fast and efficient manner ready for the correlator. This has been successfully demonstrated between Warkworth and the IBM Blade Server at AUT.

We have found KAREN to be a very useful tool for transmitting data to our international partners, and our Institute will be extending its use over the coming months. The next challenge will be to establish the IVS regular observational sessions and transmit the even larger volumes of data to the correlation centres, such as Bonn and USNO. Of future benefit to our work to stream data real-time to the correlation centres would be the ability to reserve bandwidth as a logical pipe within the KAREN bandwidth for the duration of an experiment.

## ACKNOWLEDGEMENTS

This research was supported by the Ministry for Economic Development's SKA Grant. Warkworth connectivity via KAREN was made possible by the Remote Site Connectivity Fund provided by the Ministry of Research Science and Technology (MoRST) on behalf of the New Zealand Government. We are grateful to IBM for providing the IRASR with the IBM Blade Centre in the framework of the Shared University Research (SUR) Grant.

We thank the reviewers for many useful comments that led to an improvement of the paper. In addition we wish to thank Donald Clark, Bill Choquette, John Raine and Chris Litter from REANNZ/KAREN, and Ben Yeldon from AUT's ICT Department for their support and assistance. We thank Guifre Molera of the Metsähovi Radio Observatory, Simone Bernhart of the Max Planck Institute for Radio Astronomy at Bonn and Paul Boven from JIVE, who all assisted us in establishing connectivity to their sites. We are grateful to Hiroshi Takeuchi of JAXA, Japan and all our VLBI colleagues in Australia, in particular Steven Tingay, the International Centre for Radio Astronomy Research, Curtin University of Technology and Tasso Tzioumis, CSIRO.